\documentclass[iop,apj]{emulateapj}

\usepackage{epsfig}
\usepackage{graphicx,rotating,ragged2e}
\usepackage{amsmath}

\begin{document}

\title{An Ultraviolet Spectrum of the Tidal Disruption
Flare ASASSN-14\lowercase{li}}

\author{
S.~Bradley Cenko\altaffilmark{1,2},
Antonino Cucchiara\altaffilmark{3,1},
Nathaniel Roth\altaffilmark{4},
Sylvain Veilleux\altaffilmark{5,2},
J.~Xavier Prochaska\altaffilmark{6},
Lin Yan\altaffilmark{7},
James Guillochon\altaffilmark{8},
W.~Peter Maksym\altaffilmark{9,10},
Iair Arcavi\altaffilmark{11,12},
Nathaniel R.~Butler\altaffilmark{13},
Alexei V.~Filippenko\altaffilmark{14},
Andrew S.~Fruchter\altaffilmark{3},
Suvi Gezari\altaffilmark{5},
Daniel Kasen\altaffilmark{4,14,15}
Andrew J. Levan\altaffilmark{16},
Jon M.~Miller\altaffilmark{17},
Dheeraj R.~Pasham\altaffilmark{1,2},
Enrico Ramirez-Ruiz\altaffilmark{6},
Linda E.~Strubbe\altaffilmark{18},
Nial R. Tanvir\altaffilmark{19},
and Francesco Tombesi\altaffilmark{20,5}
}

\altaffiltext{1}{Astrophysics Science Division, NASA Goddard Space Flight Center,
   Mail Code 661, Greenbelt, MD 20771, USA}
\altaffiltext{2}{Joint Space-Science Institute, University of Maryland, College 
   Park, MD 20742, USA}
\altaffiltext{3}{Space Telescope Science Institute, 3700 San Martin Drive, 
   Baltimore, MD 21218, USA}
\altaffiltext{4}{Department of Physics, University of California, Berkeley, 
   CA 94720, USA}
\altaffiltext{5}{Department of Astronomy, University of Maryland, Stadium 
   Drive, College Park, MD 20742-2421, USA}
\altaffiltext{6}{Department of Astronomy and Astrophysics, University of 
   California, Santa Cruz, CA 95064, USA}
\altaffiltext{7}{Infrared Processing and Analysis Center, California 
   Institute of Technology, Pasadena, CA 91125, USA}
\altaffiltext{8}{Harvard-Smithsonian Center for Astrophysics, The Institute 
   for Theory and Computation, 60 Garden Street, Cambridge, MA 02138, USA}
\altaffiltext{9}{Harvard-Smithsonian Center for Astrophysics, 60 Garden 
   Street, Cambridge, MA 02138, USA}
\altaffiltext{10}{Department of Physics and Astronomy, University of 
   Alabama, Tuscaloosa, AL 35487, USA}
\altaffiltext{11}{Las Cumbres Observatory Global Telescope, 6740 Cortona Dr, 
   Suite 102, Goleta, CA 93111, USA}
\altaffiltext{12}{Kavli Institute for Theoretical Physics, University of 
   California, Santa Barbara, CA 93106, USA}
\altaffiltext{13}{School of Earth \& Space Exploration, Arizona State University, 
   AZ 85287, USA}
\altaffiltext{14}{Department of Astronomy, University of California, Berkeley, 
   CA 94720-3411, USA}
\altaffiltext{15}{Nuclear Science Division, Lawrence Berkeley National 
   Laboratory, Berkeley, CA, 94720, USA}
\altaffiltext{16}{Department of Physics, University of Warwick, 
   Coventry, CV4 7AL, UK}
\altaffiltext{17}{Department of Astronomy, The University of Michigan, 1085 
   South University Avenue, Ann Arbor, Michigan 48103, USA} 
\altaffiltext{18}{Department of Physics and Astronomy \& Carl Wieman Science 
   Education Initiative, University of British Columbia, 6224 Agricultural Rd, 
   Vancouver, BC V6T 1Z1, Canada}   
\altaffiltext{19}{Department of Physics and Astronomy, University of 
   Leicester, Leicester, LE1 7RH, UK}
\altaffiltext{20}{X-ray Astrophysics Laboratory, NASA/Goddard Space 
   Flight Center, Greenbelt, Maryland 20771, USA}

   
\email{Email: brad.cenko@nasa.gov}


\shorttitle{UV Spectroscopy of ASASSN-14li}
\shortauthors{Cenko et al.}


\newcommand{\Swift}{\textit{Swift}}
\newcommand{\fermi}{\textit{Fermi}}
\newcommand{\mgii}{\ion{Mg}{2} $\lambda \lambda$2796, 2804}
\newcommand{\oii}{[\ion{O}{2} $\lambda$3727]}
\newcommand{\ip}{\textit{i$^{\prime}$}}
\newcommand{\zp}{\textit{z$^{\prime}$}}
\newcommand{\gp}{\textit{g$^{\prime}$}}
\newcommand{\rp}{\textit{r$^{\prime}$}}
\newcommand{\up}{\textit{u$^{\prime}$}}
\newcommand{\cii}{\ion{C}{2} $\lambda$1335}
\newcommand{\ciii}{\ion{C}{3}] $\lambda$1909}
\newcommand{\mgi}{\ion{Mg}{1} $\lambda$2853}
\newcommand{\civ}{\ion{C}{4} $\lambda \lambda$1548, 1551}
\newcommand{\siv}{\ion{S}{4} $\lambda \lambda$1394, 1403}
\newcommand{\nv}{\ion{N}{5} $\lambda \lambda$1239, 1243}
\newcommand{\heiiA}{\ion{He}{2} $\lambda$1640}
\newcommand{\heiiB}{\ion{He}{2} $\lambda$4686}
\newcommand{\oiii}{\ion{O}{3}] $\lambda$1663}
\newcommand{\niii}{\ion{N}{3}] $\lambda$1750}
\newcommand{\niv}{\ion{N}{4}] $\lambda$1486}

 
\begin{abstract}
We present a \textit{Hubble Space Telescope} STIS spectrum of
ASASSN-14li, the first rest-frame ultraviolet (UV) spectrum of a tidal
disruption flare (TDF).  The underlying continuum is well fit by
a blackbody with $T_{\mathrm{UV}} = 3.5 \times 10^{4}$\,K, an
order of magnitude smaller than the temperature inferred from
X-ray spectra (and significantly more precise than previous
efforts based on optical and near-UV photometry).    
Superimposed on this blue continuum, we detect three 
classes of features: narrow absorption from the Milky Way
(probably a high-velocity cloud), and narrow absorption and
broad ($\sim 2000$-8000\,km\,s$^{-1}$) emission lines 
at or near the systemic host velocity.  The absorption lines are
blueshifted with respect to the emission lines by $\Delta v = 
-(250$--400)\,km\,s$^{-1}$.  Due both to this velocity offset 
and the lack of common 
low-ionization features (\ion{Mg}{2}, \ion{Fe}{2}), we argue 
these arise from the same absorbing material responsible for the
low-velocity outflow discovered at X-ray wavelengths.  The
broad nuclear emission lines display a remarkable abundance
pattern: \ion{N}{3}], \ion{N}{4}], and \ion{He}{2} are quite 
prominent, while the common quasar emission lines of \ion{C}{3}]
and \ion{Mg}{2} are weak or entirely absent.  
Detailed modeling of this spectrum will help elucidate fundamental
questions regarding the nature of the emission processes at work
in TDFs, while future UV spectroscopy of ASASSN-14li would help 
to confirm (or refute) the previously proposed connection between TDFs and 
``N-rich'' quasars.  
\end{abstract}


\section{Introduction}
\label{sec:intro}
A star passing close to a supermassive black hole (SMBH;
$M \gtrsim 10^{6}$\,M$_{\odot}$) will be torn apart by tidal
forces \citep{h75}.  The accretion of the resulting bound stellar
debris results in a luminous transient known as a tidal 
disruption flare (TDF; \citealt{r88}).  Unlike active galactic 
nuclei (AGN), the accretion resulting
from a TDF is deterministic, and the rate that mass
first returns to the SMBH is straightforward to derive \citep{p89}.
For some systems the accretion rate is predicted to 
transition from highly super-Eddington to sub-Eddington on
a time scale of months to years (e.g., \citealt{dgn+12}).  
Furthermore, TDFs can serve
as ``sign-posts,'' indicating the presence of a SMBH in 
galaxies that are otherwise not actively accreting.  
 
\begin{figure*}[t!]
  \centerline{\includegraphics[width=18cm]{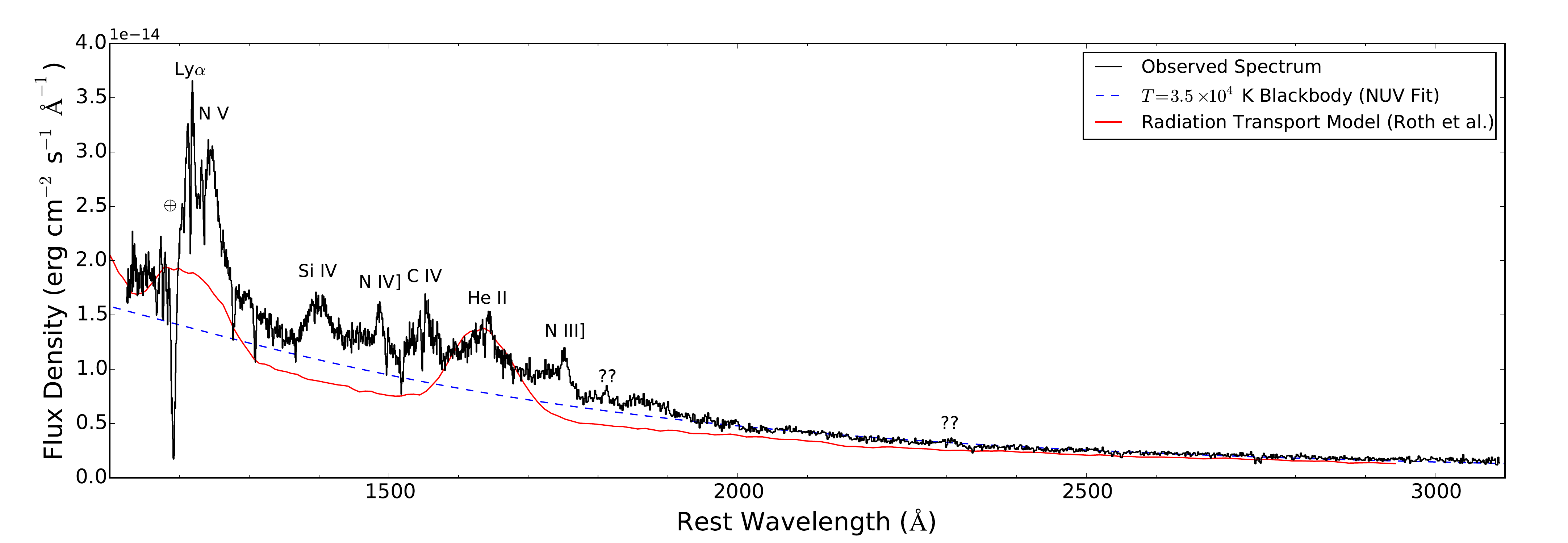}}
  \caption[]{
  UV spectrum of the TDF ASASSN-14li.  The broad emission features
  (where known) are indicated.  Overplotted is a blackbody fit
  to the (largely featureless) NUV portion of the spectrum, with
  $T_{\mathrm{BB}} = (3.50 \pm 0.06) \times 10^{4}$\,K.  Also shown
  is a radiative-transfer model from \citet{rkg+15}.  The portion 
  of the spectrum affected by geocoronal airglow is indicated with
  a circled plus sign.
  }
 \label{fig:specobs}
\end{figure*}

Thanks in part to the rapid growth in wide-field optical surveys,
the TDF discovery rate has experienced a remarkable increase in recent
years\footnote{See, e.g., \url{https://tde.space}.}.  
Optical spectra of PS1-10jh, the first TDF with extensive
real-time follow-up observations from such surveys, revealed broad 
[full width at half-maximum intensity (FWHM) $\approx 9000$\,km\,s$^{-1}$] 
\ion{He}{2} emission lines, but 
lacked detectable Balmer H emission \citep{gcr+12}.  The low H:He ratio 
has led to a vigorous debate within the community, with possible
explanations including the disruption of a H-poor star 
\citep{gcr+12,bca14,sm15}; optical-depth effects \citep{gr14}
in a radially truncated (but H-rich) broad-line region 
\citep[BLR;][]{gmr14}; complex photoionization processes within
an (H-rich) envelope surrounding the accretion disk \citep{rkg+15};
and stellar evolution (i.e., H burning) in the core of a
$\gtrsim 1$\,M$_{\odot}$ star \citep{k15}.  More recently, a larger 
sample of optically discovered TDFs has revealed a wide range
of H:He line ratios \citep{ags+14}, and still others appear to lack
emission lines altogether (e.g., \citealt{cbk+12,cbg+14}).  

Understanding the nature of this emission is critical for developing
a theoretical framework to robustly infer SMBH properties (e.g., mass)
from TDF observations.  To this end, we have undertaken a campaign
to obtain rest-frame ultraviolet (UV) spectra of TDF candidates with the 
\textit{Hubble Space Telescope (HST)}.  By analogy with
quasars, we anticipate that the strongest atomic lines 
will appear in the rest-frame UV.  Similarly, the more-distant 
events discovered in the near future by, for example, the Large Synoptic
Survey Telescope (LSST; \citealt{ita+08}) will be observed at these
rest-frame UV wavelengths.  In this {\it Letter} we present the first
spectrum obtained as part of this UV spectroscopy program, of 
the nearby TDF ASASSN-14li.

Throughout this work, we adopt a standard $\Lambda$CDM cosmology with
parameters from \citet{aaa+15b}: $H_{0} = 67.8$\,km\,s$^{-1}$\,Mpc$^{-1}$,
$\Omega_{m} = 0.308$, and $\Omega_{\Lambda} = 1 - \Omega_{m}$.  
All quoted uncertainties are 1$\sigma$ (68\%) confidence intervals 
unless otherwise noted, and UTC times are used throughout.

\section{ASASSN-14\lowercase{li} Discovery and Observations}
\label{sec:obs}
The All-Sky Automated Survey for Supernova (ASASSN; \citealt{spg+14})
first detected ASASSN-14li in $V$-band images obtained on 2014
Nov 11.65 in the nucleus of the galaxy PGC\,043234 \citep{ATEL.6777,hkp+16}.  
Pre-outburst observations from
the Sloan Digital Sky Survey (SDSS; \citealt{aaa+15}) indicate that the
host galaxy is dominated by an old stellar population ($t_{\mathrm{age}}
\approx 11$\,Gyr) with a stellar mass $\log_{10}(M_{*} / $M$_{\odot}) \approx 
9.7$ and a redshift of $z = 0.02058 \pm 0.00001$ 
\citep{cgw09}.  At distance $d \approx 90$\,Mpc, ASASSN-14li was
exquisitely observed across the electromagnetic spectrum 
\citep{hkp+16,vas+15,mkm+15,abg+15}.  We highlight the following previous results.
\begin{itemize}
\item The broadband spectral energy distribution requires
the presence of multiple emission components \citep{hkp+16,vas+15}.
The observed X-ray emission is well fit by a blackbody with 
$T_{\mathrm{X}} \approx 5.8 \times 10^{5}$\,K \citep{mkm+15};
however, extrapolating this model to optical wavelengths grossly 
underpredicts the observed flux.  A second blackbody with
$T_{\mathrm{opt}} \approx 4 \times 10^{4}$\,K
can reasonably describe the optical emission, though the 
peak is largely unconstrained \citep{hkp+16,vas+15}.
\item Evidence for a high-velocity outflow [$\Delta v \approx
(1.2$--$3.9) \times 10^{4}$\,km\,s$^{-1}$] and a low-velocity
outflow ($\Delta v = -($100--400)\,km\,s$^{-1}$) was derived
from nonthermal radio emission \citep{abg+15,vas+15} and
high-resolution X-ray spectroscopy \citep{mkm+15}, respectively.
Outflows are predicted to accompany accretion at super-Eddington 
rates (e.g., \citealt{omn+05,sq09}), though typically only with
large velocities.
\item The optical spectra of ASASSN-14li are dominated by a blue
continuum and asymmetric, broad [FWHM $\approx (1$--$2)\times 10^{4}$\,km\,s$^{-1}$]
emission lines of Balmer H and \ion{He}{2} \citep{hkp+16}.  
\end{itemize}

We obtained UV spectra of ASASSN-14li with the Space Telescope
Imaging Spectrograph (STIS; Program ID GO-13853) 
on \textit{HST} beginning at 08:40
on 2015 Jan 10 ($\Delta t = 59.7$\,d after initial detection).
STIS was deployed with two different instrumental configurations,
both utilizing the 52\arcsec $\times$ 0.2\arcsec\ aperture: the
G230L grating with the near-UV (NUV) MAMA detector, providing wavelength 
coverage of 1570--3180\,\AA\ at a resolution ($R \equiv \lambda / 
\Delta \lambda$) of $\sim 700$ (1750\,s exposure time), and the 
G140L grating with the far-UV (FUV) MAMA detector, providing wavelength 
coverage of 1150--1730\,\AA\ at $R \approx 1700$ (2888\,s 
exposure time).  

We downloaded the processed frames from the \textit{HST} archive and examined
the resulting two-dimensional spectra.  For both the FUV and NUV spectra,
the trace from ASASSN-14li is well detected and (spatially)
unresolved, so we make use of the standard pipeline product one-dimensional 
spectra for our analysis.  After a signal-to-noise ratio
(SNR)-weighted combination of the FUV and NUV frames and dereddening
for absorption in the Milky Way [$E(B-V) = 0.022$\,mag; \citealt{sf11}], 
the resulting UV spectrum of ASASSN-14li is plotted in 
Figure~\ref{fig:specobs}.  We caution that wavelengths near
1216\,\AA\ are significantly affected by geocoronal
airglow emission and will not be used for analysis here.  Upon
publication we will make the spectrum of ASASSN-14li available
via WISeREP \citep{yg12}.

\section{Analysis}
\label{sec:analysis}

\subsection{UV Continuum Emission}
\label{sec:continuum}
To estimate the continuum level, we fit the 
portion of the spectrum with $\lambda \ge 1900$\,\AA\ to a blackbody
function\footnote{We assume the host-galaxy contribution at these 
wavelengths is minimal, as indicated by pre-outburst
\textit{GALEX} photometry \citep{mkm+15}.}.  
The best-fit model, with $T_{\mathrm{BB}} = (3.50 \pm 0.06)
\times 10^{4}$\,K, is overplotted in Figure~\ref{fig:specobs}.  The
FUV flux is, as expected, somewhat underpredicted by this model, 
but it seems to capture the continuum underlying the broad FUV
emission features reasonably well.  The total bolometric luminosity
implied is $L_{\mathrm{BB}} = 2.4 \times 
10^{43}$\,erg\,s$^{-1}$, a factor of $\sim$2 larger than the 
integrated emission in the observed bandpass [$L_{\mathrm{UV}} = 
(1.15 \pm 0.06) \times 10^{43}$\,erg\,s$^{-1}$].  

The derived UV continuum temperature at this epoch is consistent
with previous estimates based on blackbody fits to broadband
optical and near-UV photometry \citep{hkp+16,vas+15}.  As noted
by these authors, a single blackbody cannot simultaneous account
for the luminous optical/UV and X-ray emission.  

\begin{figure*}[t!]
  \centerline{\includegraphics[width=18cm]{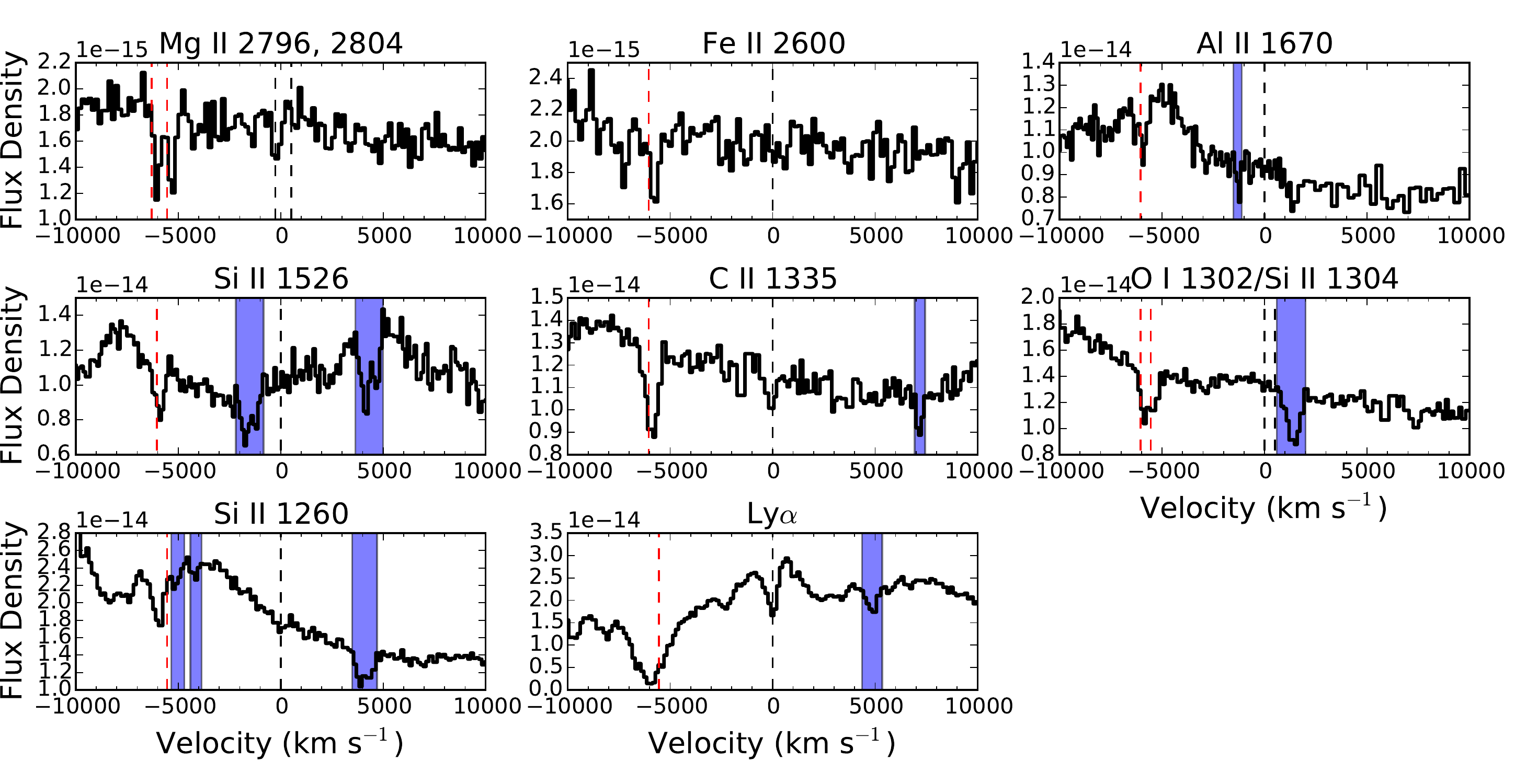}}
  \caption[]{
  Low-ionization (ionization energy $< 13.6$\,eV) lines in the 
  spectrum of ASASSN-14li.  For each subplot, we have normalized 
  the velocity to the SDSS redshift of the host galaxy, $z = 0.02058$.  
  Common absorption features 
  are detected from the Milky Way ISM (red vertical dashed lines for 
  $v_{\mathrm{MW}} = 0$), with a significant velocity offset.  Regions
  contaminated by features from other transitions are shaded blue.
  }
 \label{fig:lowion}
\end{figure*}

In Figure~\ref{fig:specobs}, we also
overplot the results of a radiative-transfer simulation from 
\citet{rkg+15}.  This particular geometry
consists of an outer envelope of 0.125\,M$_{\odot}$ of material
extending out to $5 \times 10^{14}$\,cm (a factor of $\sim 3$ larger
than the radius inferred from our blackbody fits above), with a 
density profile of $\rho \propto r^{-2}$.  The envelope is 
illuminated by an underlying continuum with a photospheric radius 
of $10^{14}$\,cm and a bolometric luminosity of $10^{45}$\,erg\,s$^{-1}$.  
The resulting output spectrum is not a formal ``fit'' to the data, and 
furthermore only contains atomic species of H, He, and O at solar abundances.  
However, it is illustrative that the observed
UV continuum slope is quite well reproduced with such a geometry.

\subsection{Absorption and Emission Features}
\label{sec:features}
We fit emission and absorption 
features to a Gaussian model, allowing the central wavelength, line width,
and equivalent width to vary as free parameters.  The local continuum level
was estimated from nearby wavelength bins.  The results of this
analysis are displayed in Table~\ref{tab:features}.  We identify three
distinct classes of features: narrow absorption from the Milky Way, 
and narrow absorption and broad emission at or near the host redshift.

\subsubsection{Milky Way Absorption}
\label{sec:mwabs}
We detect a series of narrow (FWHM $\approx 500$\,km\,s$^{-1}$) 
absorption features from standard metal transitions in the 
interstellar medium (ISM) at wavelengths near their
rest values.  Given their proximity to rest wavelengths and the lack
of additional intervening material, we associate these features with
an absorber in or near the Milky Way Galaxy.   We calculate a weighted 
average redshift for the absorber of $z = 0.00064 \pm 0.00006$, or a 
velocity relative to the heliocentric reference frame of $v = 190 \pm 
20$\,km\,s$^{-1}$.  For the Galactic coordinates of ASASSN-14li 
($l = 298.28^\circ$, $b = 80.62^\circ$), this results in a comparable velocity 
in the Local Standard of Rest. This velocity is indicative of a 
high-velocity cloud (HVC; \citealt{ww97}).  

\subsubsection{Absorption at Host Redshift}
\label{sec:hostabs}
\begin{figure*}[t!]
  \centerline{\includegraphics[width=18cm]{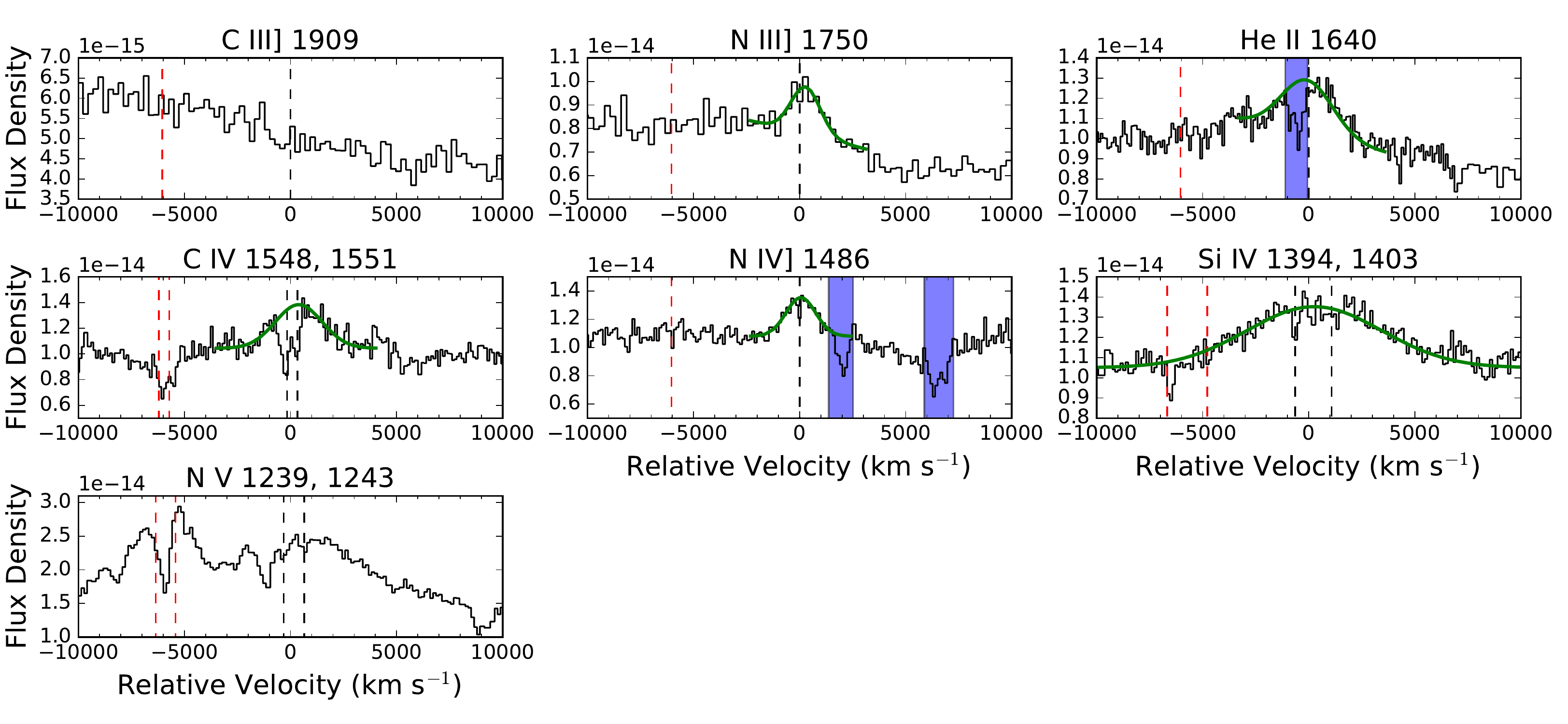}}
  \caption[]{
  High-ionization (ionization energy $> 13.6$\,eV) lines in the 
  spectrum of ASASSN-14li.  The plot scheme is the same as in 
  Figure~\ref{fig:lowion}.  Best-fit Gaussian emission models
  are plotted in green.  
  }
 \label{fig:highion}
\end{figure*}

While the absorption features from the Milky Way generally resemble
those observed in the ISM of high-redshift galaxies 
(e.g., \citealt{wgp05,fjp+09}), the narrow absorption features observed from 
the host galaxy of ASASSN-14li are markedly different.  With the 
notable exception of weak \cii, nearly 
all transitions from low-ionization metal states are absent 
(Figure~\ref{fig:lowion})\footnote{The feature at 
$\lambda_{\mathrm{obs}} = 2854$\,\AA\ could correspond either to 
\mgi\ from the Milky Way or \ion{Mg}{2}
$\lambda$ 2796 from the TDF host.  Given the lack of corresponding
\ion{Mg}{2} $\lambda$ 2803 at the host redshift, we associate this 
with the Milky Way absorber.}.  Similarly, Ly$\alpha$ is extremely
weak: assuming the gas is optically thin, we derive a lower limit to
the column density of $\log N($cm$^{-2}) \ge 14.2 \pm 0.2$.  At first glance, the weak
absorption is not entirely surprising, given the old stellar
population observed in the host galaxy. But standard high-ionization 
absorption lines, such as \civ, \siv, and \nv\ are well detected (Figure~\ref{fig:highion}).  
We measure a weighted average for the absorber redshift of 
$z = 0.02044 \pm 0.00006$, marginally below the value measured in 
the (quiescent) host.      

\subsubsection{Emission at Host Redshift}
\label{sec:hostem}
A number of broad emission features are also apparent, 
including \civ, \siv, \nv, and Ly$\alpha$.  We further identify the
feature at $\lambda_{\mathrm{obs}} = 1674$\,\AA\ as \heiiA, 
given the strong \heiiB\  
emission observed in the optical spectra of ASASSN-14li
\citep{hkp+16}.  This line was predicted to appear in the 
UV spectra of TDFs by \citet{sm15}, but with a P-Cygni profile
not readily identified here.  It also appears prominently in
the radiative-transfer simulations of \citet{rkg+15} 
(Figure~\ref{fig:specobs}).\footnote{We note that \citet{rkg+15}
adopted a Gaussian line profile with Doppler velocity of
$10^{4}$\,km\,s$^{-1}$ for their spectra; this profile and line
width were not an output of the simulation.}  This \ion{He}{2} 
line may be blended with \oiii, but the low SNR
precludes firm conclusions. 

We associate two additional features with \niii\ and \niv\ (see also 
\citealt{k15}).  We are unable to identify two remaining 
lines, both relatively weak, at $\lambda_{\mathrm{host}} \approx
2303$\,\AA\ and $\lambda_{\mathrm{host}} \approx 
1812$\,\AA.\footnote{The latter line appears in the 
composite N-rich QSO spectrum presented in \citet{jfv08} and
may result from an \ion{Si}{2} blend; see their Figure 4.}    
Equally of interest are common quasar absorption features that
are entirely lacking from the UV spectrum of ASASSN-14li:
\mgii\ and \ciii\ (Figure~\ref{fig:comp}).  

After manually excising narrow absorption features in the regions
of interest, we fit the emission lines to a Gaussian model,
as described above.  
The width of the emission features varies considerably, 
from FWHM $\approx 7700$\,km\,s$^{-1}$ for 
\siv\ to 1700\,km\,s$^{-1}$ for 
\niv\ (corresponding to distances of $\sim 7$--150\,AU 
for virialized gas orbiting a 
$10^{6}$\,M$_{\odot}$ SMBH; \citealt{vas+15,hkp+16}).  
In fact, aside from one of the 
unidentified features, the two N transitions are significantly 
more narrow than the remainder of the lines.  
Even the
largest velocities measured here are substantially smaller
than those measured from optical lines 
[FWHM $\approx (1$--$2)\times 10^{4}$\,km\,s$^{-1}$; 
\citealt{hkp+16}].

For features where both narrow absorption and broad emission
are detected, the absorption features are blueshifted:
$\Delta v = -450 \pm 100$\,km\,s$^{-1}$ for \civ, and 
$\Delta v = -250 \pm 110$\,km\,s$^{-1}$ for \siv.  

Finally, we note that we did not attempt to fit either the
broad Ly$\alpha$ or \nv\ host emission features, owing to both their clear 
blending and the large number of narrow absorption features
at these wavelengths.  
The absorption feature blueward of Ly$\alpha$ is from \ion{H}{1}
gas in the HVC; none of the lines shows evidence for a P-Cygni
profile.

\section{Discussion}
\label{sec:discussion}
\begin{figure*}[t!]
  \centerline{\includegraphics[width=18cm]{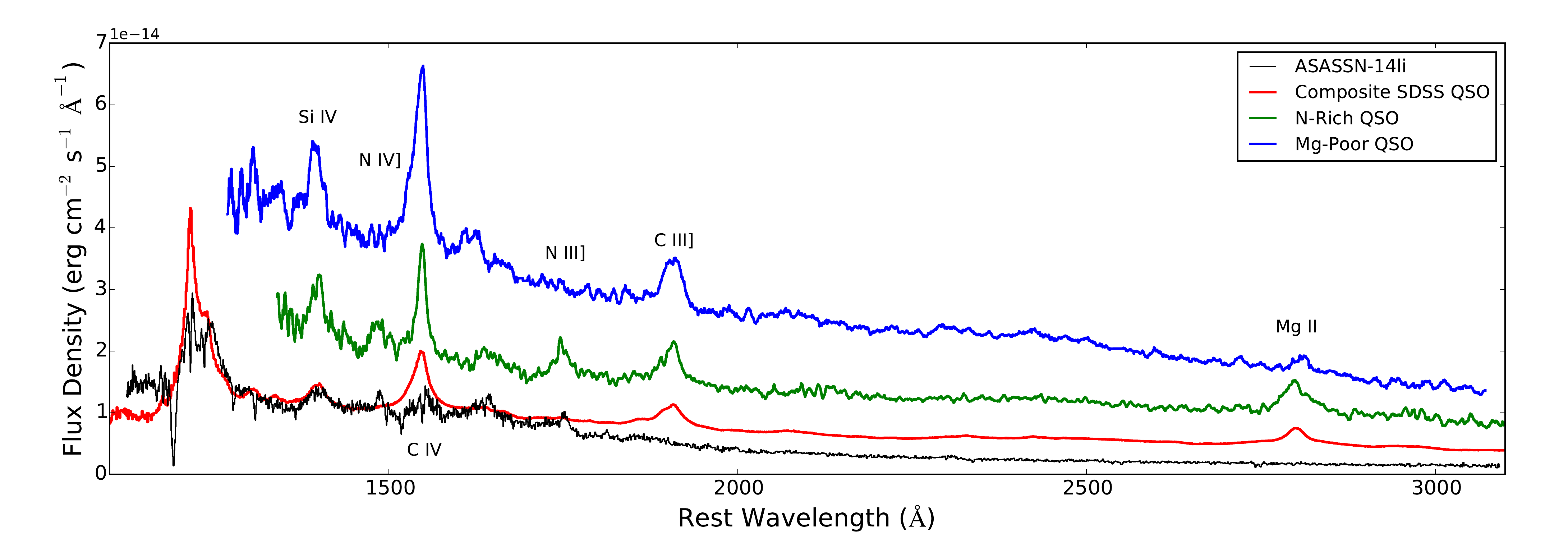}}
  \caption[]{
  The UV spectrum of ASASSN-14li, compared with the composite
  QSO spectrum from SDSS (red; \citealt{vrb+01}), a ``N-rich''
  QSO (SDSS\,J125414.27+024117.5), and an ``Mg-poor'' QSO
  (SDSS\,J003922.44+005951.7).
  }
 \label{fig:comp}
\end{figure*}

As a starting place for discussion, the most
natural point of comparison is with quasars (QSOs), given that both
are powered by accretion onto a SMBH.  In Figure~\ref{fig:comp} we
overplot the composite QSO spectrum from SDSS \citep{vrb+01}.  Even
compared with a QSO, the continuum emission from ASASSN-14li is 
significantly bluer.  Together with the luminous, thermal X-ray
emission, the nature of the continuum in ASASSN-14li appears to be
significantly different from that observed in typical QSOs.  

In terms of line features, our first task is to understand the source
of the emitting and absorbing material.  The lack of common
low-ionization absorption features, together with the old host 
stellar population, indicate that the absorber is unlikely to arise
in the cold host ISM.  Given that nearly every $L_{*}$ galaxy 
exhibits a circumgalactic medium (GGM) consisting of strong 
Ly$\alpha$ absorption and (frequently) low-ionization absorption 
\citep{pwc+11,ttw+12}, there is a reasonable
probability that gas in the halo of the host galaxy 
contributes to the observed absorption.  However, the detection of
\nv\ is difficult to account for in such a CGM environment
(Werk et al., in preparation).  

The host galaxy does show evidence for
a pre-outburst AGN, and so the absorber could result from pre-existing
gas in (or possibly expelled from; \citealt{vas+15}) the nuclear region.
Alternatively, the absorber may result from the (bound) debris of
the disrupted star.  We argue this is the simplest explanation,
as the same material responsible for the X-ray outflow 
\citep{mkm+15} could also produce UV absorption (e.g., 
\citealt{ckb+99,k06}).  The comparable blueshift (with respect
to the broad emission lines) observed in the UV and X-rays further
supports this conclusion.

The emission lines present in ASASSN-14li are to first order analogous
to a BLR in a QSO: clearly there is fast-moving 
($\gtrsim 5 \times 10^{3}$\,km\,s$^{-1}$) material photoionized (or collisionally
excited) by the blue continuum.  However, closer examination reveals 
an extremely unusual abundance pattern.  Strong
\niv\ and \niii\ are extremely
rare in QSOs, with only $\sim1$\% of SDSS systems exhibiting such
features \citep{jfv08}.  The lack of \ciii\ and
\mgii\ emission is particularly striking.

We have searched through the SDSS QSO database to identify other
examples of systems that either (a) exhibit strong \niv\ and 
\niii\ emission, and/or (b) display 
large \civ\ equivalent widths, but
weak or nonexistent \mgii\ emission.  
As noted by \citet{k15}, the former group has been previously 
identified as the ``N-rich'' QSOs \citep{o80,bo04,bho04}.  \citet{k15}
argues that these N-rich QSOs are in fact the tidal disruption of
a $\gtrsim 1$\,M$_{\odot}$ star, for which the high N abundance
results from significant CNO processing in the stellar core
(predisruption).

While the similarity between N-rich QSOs and ASASSN-14li is indeed
intriguing, we note several important caveats.  First, we examined
all the N-rich systems identified by \citet{jfv08}, and every one 
(in the appropriate redshift range) exhibited detectable 
\ciii\ and \mgii\
emission lines.  Mg should be unaffected by the CNO cycle.  The absence of 
\ion{Mg}{2} could be the result of photoionization 
from the extremely hot continuum: given that the ionization energy
of \ion{Mg}{3} is 80.1\,eV (comparable to the ionization energy
of \ion{N}{4} of 77.5\,eV), there are clearly many photons
capable of further electron stripping.  In this case, the
absence of \ion{Mg}{2} may be \textit{transient} in nature,
as the underlying continuum temperature must cool eventually.

In addition, only a small fraction ($\lesssim 10\%$) of the N-rich
QSOs exhibit narrow, blueshifted absorption lines
as were observed for ASASSN-14li.  Again, however, if related to
accretion at super-Eddington rates, these absorption features could
also be transient in nature.  Future spectra of ASASSN-14li, when
the accretion rate has dropped and the continuum cooled, are a 
critical test of this explanation.


We also plot in Figure~\ref{fig:comp} an example QSO spectrum with
well-detected \civ, but
weak or nonexistent \mgii\ emission.
Of the 78,223 QSOs in the catalog of \citet{srs+11} with well-detected
\civ\ and $z < 2$, only 269
(0.3\%) lack detectable \mgii\ ($W_{r} < 
3\sigma$).  We visually inspected these
269 ``Mg-poor'' QSOs, and in nearly all cases the lack of 
\mgii\ is due to decreased
sensitivity at the relevant (observed) wavelength.  All these
sources have well-detected \ciii\ lines.  

Detailed photoionization modeling may help to shed further light
on the nature of the emitting gas.  For example, the lack of 
\ciii\ could be explained if the gas were above
the critical density of $n_{\mathrm{crit}} = 10^{9.5}$\,cm$^{-3}$
\citep{o89}; constraints from the X-ray spectrum place a lower
limit of $n \gtrsim 2 \times 10^{9}$\,cm$^{-3}$.  Similarly, if
the narrow width of the (semi-forbidden) N lines were caused by
collisional de-excitation (and not, say, distance from the source),
this may also constrain the density of the emitting gas.  However,
we caution that assumptions underlying standard AGN photoionization
tools (e.g., CLOUDY; \citealt{fpv+13}) may not hold for TDFs
\citep{rkg+15}.
 
To the extent that ASASSN-14li
is representative of the broader TDF population, the UV
spectrum is extremely promising for the \textit{detection}
of high-redshift events by future wide-field optical surveys.
With the blue continuum (and no evidence for dust), the
resulting negative K-correction will greatly enhance 
detectability: ASASSN-14li, for example, would be easily 
detectable by LSST out to $z \approx 1$ ($m_{g} \approx
25.0$\,mag at $d_{\mathrm{L}} = 6.8$\,Gpc).  Furthermore, the extremely
blue continuum should clearly distinguish TDFs from other 
classes of transients \citep{vfg+11}.

However, based on this result, it is clear
that we have yet to reach a complete picture of the process
by which the emission is generated following the tidal
disruption process.  The simplest analytic models (e.g., 
\citealt{u99,sq09,lr11}), which assume rapid circularization of the
bound debris, may miss fundamental physics governing the
observed signal \citep{skc+15,psk+15,gr15}.  As is clear 
from our spectrum of ASASSN-14li, the line formation is a complex 
process, and QSOs are an imperfect analog.  Without improved 
understanding of how (and where) the emission is generated, it will be 
quite challenging to utilize TDFs as probes of distant
SMBHs.

Finally, the path forward (observationally) is quite clear.
Time-resolved spectra  should solve many
of the puzzles presented here.  
A reverberation mapping campaign (e.g., \citealt{p93})
would be an incredibly powerful discriminant for future events.
Furthermore, additional nearby examples are necessary to determine if ASASSN-14li 
is indeed a fair representation of the broader TDF population,
or if something unique about ASASSN-14li (nature of the 
disrupted star, orbit, geometry, etc.) gives rise to both
luminous X-ray and optical/UV emission.


\acknowledgments
We thank R.~Chornock, M.~Eracleous, P.~Hall, and C.~Kochanek for valuable
discussions, and the {\it HST} staff for the prompt scheduling of 
these ToO observations.  S.B.C. acknowledges the Aspen Center 
for Physics and NSF Grant $\#1066293$ for hospitality.  
A.V.F.'s research was funded by NSF grant AST-1211916, 
the TABASGO Foundation, and the Christopher R. Redlich Fund. 

Based on observations made with the NASA/ESA {\it Hubble Space Telescope}, obtained 
from the Data Archive at the Space Telescope Science Institute, which is 
operated by the Association of Universities for Research in Astronomy, Inc., 
under NASA contract NAS 5-26555.
 
\vspace{0.5cm}

{\it Facilities:} \facility{HST (STIS)}

\bibliographystyle{apj}

\clearpage

\begin{deluxetable*}{cccccccc}
\tabletypesize{\scriptsize}
\tablecaption{ASASSN-14li Absorption and Emission Features}
\tablewidth{0pt}
\tablehead{
\colhead{$\lambda_{\mathrm{obs}}$} & \colhead{Identification} &
 \colhead{$\lambda_{0}$} &
 \colhead{$W_{\mathrm{r}}$\tablenotemark{a}} & \colhead{FWHM} &
 \colhead{Line Flux} & \colhead{$z$} & \\
\colhead{(\AA)} & & \colhead{(\AA)} & \colhead{(\AA)} &
 \colhead{(\AA)} & \colhead{($10^{-14}$\,erg\,cm$^{-2}$\,s$^{-1}$)} 
}
\startdata
2911.27 & \ion{Mg}{1} & $2852.96$ & $< 1.0$ & 3.0 & $\cdots$ & 0.02044 \\

2860.83 & \ion{Mg}{2} & $2803.53$ & $< 0.65$ & 3.0 & $\cdots$ & 0.02044 \\

2856.3 & \ion{Mg}{2} & 2798.75 & $> -2.8$ & 15.0 & $< 0.8$ & 0.02058 \\

$2854.35 \pm 0.42$ & \ion{Mg}{1} & $2852.96$ & $0.95 \pm 0.36$ & 
 $4.0 \pm 1.4$ & $\cdots$ & $0.00049 \pm 0.00016$ \\
  
$2805.70 \pm 0.24$ & \ion{Mg}{2} & $2803.53$ & $1.49 \pm 0.24$ &
 $3.8 \pm 0.6$ & $\cdots$ & $0.00077 \pm 0.00009$ \\
$2799.04 \pm 0.21$ & \ion{Mg}{2} & $2796.35$ & $1.33 \pm 0.22$ & 
 $3.2 \pm 0.5$ & $\cdots$ & $0.00096 \pm 0.00008$ \\

2653.32 & \ion{Fe}{2} & $2600.17$ & $< 0.77$ & 3.0 & $\cdots$ & 0.02044 \\

$2602.67 \pm 0.33$ & \ion{Fe}{2} & $2600.17$ & $0.79 \pm 0.20$ & 
 $3.3 \pm 0.9$ & $\cdots$ & $0.00096 \pm 0.00013$ \\

2431.47 & \ion{Fe}{2} & $2382.77$ & $< 0.61$ & 3.0 & $\cdots$ & 0.02044 \\

2392.13 & \ion{Fe}{2} & $2344.21$ & $< 0.64$ & 3.0 & $\cdots$ & 0.02044 \\

$2385.21 \pm 0.35$ & \ion{Fe}{2} & $2382.77$ & $0.69 \pm 0.24$ &
 $3.0 \pm 1.1$ & $\cdots$ & $0.00102 \pm 0.00015$ \\
 
$2350.0 \pm 1.6$ & ??? & $\cdots$ & $-2.9 \pm 0.8$ & $20.7 \pm 6.0$ & 
 $0.76 \pm 0.27$ &  $\cdots$ \\

$2346.78 \pm 0.20$ & \ion{Fe}{2} & $2344.21$ & $0.59 \pm 0.19$ &
 $2.0 \pm 0.6$ & $\cdots$ & $0.00110 \pm 0.00009$ \\

1948.0 & \ion{C}{3} & 1908.73 & $> -1.5$ & 15.0 & $< 0.5$ & 0.02058 \\

$1849.3 \pm 0.9$ & ??? & $\cdots$ & $-1.1 \pm 0.3$ & $5.9 \pm 2.6$ & 
 $0.62 \pm 0.20$ & $\cdots$ \\
 
$1788.0 \pm 0.7$ & \ion{N}{3}] & 1750.26 & $-2.7 \pm 0.6$ & $9.7 \pm 2.2$ &
 $2.1 \pm 0.6$ & $0.02156 \pm 0.00040$ \\

1704.94 & \ion{Al}{2} & $1670.79$ & $< 0.33$ & 3.0 & $\cdots$ & 0.02044 \\

$1673.7 \pm 0.6$ & \ion{He}{2} & 1640.42 & $-4.5 \pm 0.6$ & $16.0 \pm 1.6$ &
 $4.6 \pm 0.5$ & $0.02028 \pm 0.00037$ \\

$1671.58 \pm 0.22$ & \ion{Al}{2} & $1670.79$ & $0.40 \pm 0.14$ &
 $1.7 \pm 0.6$ & $\cdots$ & $0.00047 \pm 0.00013$ \\

$1583.0 \pm 0.5$ & \ion{C}{4} & 1549.06 & $-4.8 \pm 0.7$ & $13.7 \pm 1.0$ &
 $5.0 \pm 0.8$ & $0.02191 \pm 0.00032$ \\

$1582.60 \pm 0.14$ & \ion{C}{4} & $1550.77$ & $0.52 \pm 0.12$ &
 $1.5 \pm 0.4$ & $\cdots$ & $0.02053 \pm 0.00009$ \\
$1579.63 \pm 0.15$ & \ion{C}{4} & $1548.20$ & $1.24 \pm 0.22$ &
 $3.0 \pm 0.5$ & $\cdots$ & $0.02030 \pm 0.00009$ \\

1557.92 & \ion{Si}{2} & $1526.71$ & $< 0.25$ & 3.0 & $\cdots$ & 0.02044 \\

$1551.93 \pm 0.23$ & \ion{C}{4} & $1550.77$ & $0.48 \pm 0.17$ &
 $1.6 \pm 0.6$ & $\cdots$ & $0.00075 \pm 0.00015$ \\
$1549.16 \pm 0.21$ & \ion{C}{4} & $1548.20$ & $1.07 \pm 0.22$ & 
 $2.8 \pm 0.6$ & $\cdots$ & $0.00062 \pm 0.00014$ \\

$1527.37 \pm 0.14$ & \ion{Si}{2} & $1526.71$ & $0.90 \pm 0.16$ &
 $2.8 \pm 0.5$ & $\cdots$ & $0.00043 \pm 0.00009$ \\
 
$1517.3 \pm 0.3$ & \ion{N}{4}] & 1486.50 & $-2.3 \pm 0.4$ & $8.4 \pm 1.1$ &
 $2.4 \pm 0.4$ & $0.02072 \pm 0.00020$ \\
 
1431.44 & \ion{Si}{4} & 1402.77 & $< 0.21$ & 3.0 & $\cdots$ & 0.02044 \\
 
$1426.7 \pm 0.5$ & \ion{Si}{4} & 1396.76 & $-10.9 \pm 1.1$ & $35.7 \pm 2.0$ &
 $11.5 \pm 1.0$ & $0.02144 \pm 0.00036$ \\
 
$1422.48 \pm 0.17$ & \ion{Si}{4} & $1393.76$ & $0.21 \pm 0.07$ &
 $1.3 \pm 0.5$ & $\cdots$ & $0.02061 \pm 0.00012$ \\

$1394.53 \pm 0.13$ & \ion{Si}{4} & $1393.76$ & $0.20 \pm 0.06$ &
 $0.8 \pm 0.3$ & $\cdots$ & $0.00055 \pm 0.00009$ \\
 
$1361.66 \pm 0.19$ & \ion{C}{2} & $1334.53$ & $0.35 \pm 0.11$ &
 $2.5 \pm 0.6$ & $\cdots$ & $0.02033 \pm 0.00014$ \\

$1335.27 \pm 0.08$ & \ion{C}{2} & $1334.53$ & $1.26 \pm 0.14$ &
 $3.4 \pm 0.3$ & $\cdots$ & $0.00055 \pm 0.00006$ \\

1331.03 & \ion{Si}{2} & $1304.37$ & $< 0.11$ & 3.0 & $\cdots$ & 0.02044 \\

1328.79 & \ion{O}{1} & $1302.17$ & $< 0.12$ & 3.0 & $\cdots$ & 0.02044 \\

$1305.22 \pm 0.14$ & \ion{Si}{2} & $1304.37$ & $0.39 \pm 0.07$ &
 $1.4 \pm 0.3$ & $\cdots$ & $0.00065 \pm 0.00011$ \\

$1302.98 \pm 0.10$ & \ion{O}{1} & $1302.17$ & $0.52 \pm 0.07$ &
 $1.6 \pm 0.3$ & $\cdots$ & $0.00062 \pm 0.00008$ \\

1286.18 & \ion{Si}{2} & $1260.42$ & $<0.35$ & 3.0 & $\cdots$ & 0.02044 \\

$1268.60 \pm 0.22$ & \ion{N}{5} & $1242.80$ & $0.30 \pm 0.13$ &
 $0.7 \pm 0.5$ & $\cdots$ & $0.02076 \pm 0.00018$ \\
 
$1264.40 \pm 0.25$ & \ion{N}{5} & $1238.82$ & $0.52 \pm 0.14$ & 
 $1.2 \pm 0.5$ & $\cdots$ & $0.02065 \pm 0.00020$ \\

$1260.74 \pm 0.09$ & \ion{Si}{2} & $1260.42$ & $0.59 \pm 0.08$ &
 $2.3 \pm 0.3$ & $\cdots$ & $0.00025 \pm 0.00007$ \\

$1240.49 \pm 0.04$ & Ly$\alpha$ & $1215.67$ & $0.83 \pm 0.04$ &
 $2.0 \pm 0.1$ & $\cdots$ & $0.02042 \pm 0.00003$ 
\enddata
\tablenotetext{a}{For features associated with the host galaxy of
ASASSN-14li, we adopt the spectroscopic host redshift from SDSS, 
$z = 0.02058$, to convert equivalent widths from observed to 
rest-frame quantities.  Equivalent width is defined such that 
positive values imply absorption features.  3$\sigma$ upper limits
are provided for nondetections.}
\label{tab:features}
\end{deluxetable*}


\end{document}